# Transport of Massless Dirac Fermions in Non-topological Type Edge States

Yu.I. Latyshev[1], A.P. Orlov[1], V.A. Volkov[1,2,*], V.V. Enaldiev[1], I.V. Zagorodnev[1],

O.F. Vyvenko[3], Yu.V. Petrov[3], P. Monceau[4,5,6]

[1]*Kotelnikov Institute of Radio-engineering and Electronics of RAS, Mokhovaya 11-7, 125009 Moscow, Russia,*
[2]*Moscow Institute of Physics and Technology, Institutskii per. 9, Dolgoprudny, 141700 Moscow region, Russia,*
[3]*IRC for Nanotechnology of St. Petersburg State University, Uljanovskaya 1, Petrodvorets, 198504 St. Petersburg, Russia,*
[4]*Univ. Grenoble - Alpes, Inst. Neel, F38042 Grenoble, France,*
[5]*CNRS, Int. Neel, F38042 Grenoble, France,*
[6]*Laboratoire National des Champs Magnétiques Intenses, 25 rue des Martyrs, BP 166, 38042 Grenoble, Cedex 9, France*

[*]Correspondence to: volkov.v.a@gmail.com

**Prof. Yuri Latyshev passed away at Moscow on 10 June 2014.
The present publication is devoted to his memory.**

## Abstract

There are two types of intrinsic surface states in solids. The first type is formed on the surface of topological insulators. Recently, transport of massless Dirac fermions in the band of "topological" states has been demonstrated. States of the second type were predicted by Tamm and Shockley long ago. They do not have a topological background and are therefore strongly dependent on the properties of the surface. We study the problem of the conductivity of Tamm-Shockley edge states through direct transport experiments. Aharonov-Bohm magneto-oscillations of resistance are found on graphene samples that contain a single nanohole. The effect is explained by the conductivity of the massless Dirac fermions in the edge states cycling around the nanohole. The results demonstrate the deep connection between topological and non-topological edge states in 2D systems of massless Dirac fermions.

## Introduction

Nature is arranged such that electronic surface states (SSs) or interface states always appear on the surfaces of real crystals. Usually, they are induced by defects in the crystal surface, impurities, or contaminants and have a disordered character. Such "extrinsic" SSs lead to a finite density of states in the forbidden band of the crystal, which disturbs the functioning of many solid-state devices [1].

In principle, there may be other SSs that are not related to defects on the surface ("intrinsic" SSs). Since the pioneering works by Tamm [2] and Shockley [3], theoretical models predict that breaking of the crystal periodic potential at the surface can lead to the appearance of a two-dimensional (2D) band of conducting electronic states near the surface [4]. These SSs are sometimes called Tamm states, Shockley states, or Tamm-Shockley states. Usually, they are detected using local methods (such as STM and ARPES) on atomically clean surfaces of a number of metals and semiconductors in ultrahigh vacuum. (In addition to electronic SSs, there is an optical analogue on the surface of photonic crystals [5, 6]). However, on real interfaces, such states typically do not exist (we do not consider the special SSs that appear only in a magnetic field and are associated with skipping orbits). General criteria for the existence of Tamm-Shockley states have not been determined. What is clear is that these states are associated with both the features of the bulk band structure and the details of the electronic structure of the surface on the atomic scale. The important feature of intrinsic SSs is their ability to conduct electrical current. The high conductivity of SSs can lead to qualitatively new physical effects. Therefore, studies of their transport properties are currently of great interest. Other species of intrinsic SSs are related to the physics of topological insulators (TIs). In recent years, TI studies have been the fastest-growing area of modern physics [7, 8]. On the surface of a number of crystals (such as $Bi_2Se_3$) that have modified Dirac band structures, a 2D surface band of massless Dirac fermions (DFs) with a conical dispersion is formed. The cause of these bands is the existence of a special topological invariant, which is determined solely by the peculiarities of the bulk band structure of the TI [9]. An intriguing feature of these SSs is their transport properties: they are protected against backscattering by time-reversal symmetry, which results in their high conductance of electric current. The energy spectrum of these SSs has been investigated foremost by ARPES [10, 11, 12]. Moreover, manifestations of these SSs have also been found in transport measurements. In [13,14], an Aharonov-Bohm (AB) -type effect in the resistance of a TI -- $Bi_2Se_3$ nanowire in a longitudinal magnetic field was observed. The possibility of observing AB magneto-oscillations in samples with non-ring geometries is related to the existence of conducting SSs, encircling the nanowire cross section [15].

Thirty years ago, it became clear that narrow-gap semiconductors and semimetals with relativistic band structures (such as Bi, BiSb, and PbTe) are convenient objects for the study of Tamm-Shockley SSs. The charge carriers in these crystals are now referred to as massive DFs. For these crystals, the possibility of the formation of Tamm-Shockley-type SSs with a conical massless Dirac spectrum on the surfaces of a certain class was predicted [16]. In the model of "inversion heterojunction" [17, 18], the aforementioned specific class of surface can be attributed to the near-surface inversion mass of the bulk DFs (see also [19]). We emphasise that in this case, the bulk band structure does not exhibit the topological invariant.

Thus, at present, theory predicts the existence of two types of intrinsic SSs that are filled with massless DFs with conical spectra. The first type of SS is referred to as a topological SS. The latter is a type of Tamm-Shockley state that is referred to hereafter as a Tamm-Dirac (TD) state. The first type of SS has been detected in transport measurements, whereas the second has not. In 2D electron systems, edge states (ESs) of topological or TD type are analogous to the surface states. The present work is aimed at the detection of ESs of TD type through direct transport experiments.

Two graphene systems were selected as convenient objects of study, graphene on silicon oxide and graphene-on-graphite. Ideal graphene realises a system of 2D massless DFs that has an energy spectrum of the conical type and is doubly degenerate in the valley quantum number [20]. The theory of ESs in graphene is the subject of several theoretical works [21, 22]. In fact, these works explored ESs of the TD type, even if this fact was not explicitly stated. General graphene half-planes with translational invariant edges were considered in [23, 24]. It was predicted that the TD states spectrum in the reduced valleys scheme consists of a pair of rays that emanate from the Dirac point; see below for details. This spectrum is somewhat reminiscent of the spectrum of the topological edge states in a 2D TI of CdTe-CdHgTe type, although graphene is not a TI. It can be assumed that this similarity is not accidental and that the transport properties of these different edge states will also be similar. This assumption is the basic motivation of this work.

## Experiment

Some indications of the existence of ESs in graphene have been demonstrated using the local STM technique [25, 26]; see also the Conclusions. However, until recently, there was no direct experiment regarding their contribution to transport measurements. In the present paper, we probe TD-type ESs by measuring the Aharonov-Bohm (AB) effect. We consider that if the edge states--are conductive in nature, the AB effect will appear as resistance magneto-oscillations of graphene samples (graphene and graphene-on-graphite) that contain a single nanohole in a perpendicular magnetic field $H$. The phase and spectrum of the edge DFs that circulate around the hole will be controlled by the magnetic flux through the hole. This effect can, in principle, lead to an *H*-periodic contribution to the magnetoresistance of the sample.

Nevertheless, until recently, the AB effect in graphene has been observed only for ring-shaped samples [27-35] or on a lattice of nanoholes [36], in which the effect of a single hole was masked by the presence of the lattice and the effects of the fictitious magnetic field that is associated with elastic deformations [37]. Here, we observe for the first time the AB effect on graphene and graphite structures with a single nanohole. The latter permeates the entire structure.

We use nano-thin graphite samples to avoid shunting of the surface graphene layers by the bulk. On the surface of graphite one often finds graphene flakes weakly bound to the surface where interlayer coupling is strongly suppressed. Several papers have noted the significant contribution of the surface graphene layer to the quantum oscillations of thin graphite. Thus, recent STM [38], cyclotron resonance [39] and Raman spectroscopy [40, 41] experiments have demonstrated that the surface layer of graphite is often represented as a graphene layer of an exceptional quality ("graphene-on-graphite"). Therefore, we believe that the bulk of thin film of graphite contributes only to the background resistance, but oscillating part of the magnetoresistance is due to massless DFs in the top graphene layer. It is in agreement with the following observation: the period of these oscillations does not depend on the thickness of the graphite samples.

Large-area flakes of natural graphite with thicknesses of as little as 30 nm were cleaved from large-crystal graphite using adhesive tape, and the adhesive layer was dissolved in acetone. In the second stage, the crystal was thinned using soft plasma etching down to the atomic thickness of approximately 1 nm [42]. Scanning Raman spectroscopy indicated high uniformity of the thinned crystals over the lateral size of hundreds of microns. For comparison, we also used commercial graphene samples.

The samples were processed to form a Hall bar geometry. Nanoholes were introduced using two independent techniques, (1) the FIB technique with Ga-ions and (2) a helium ion microscope. We used the FIB technique to fabricate small holes with diameters of as little as *d*=35 nm (Fig. 1a). To produce the smallest hole, with diameter *d*=20 nm, we used the helium ion microscope (Figs. 1b and 1c).

Shubnikov-de Haas oscillations were clearly observed in the thin graphite samples in weak fields. They exhibited inversed-field periodicity with a period of 0.2 T$^{-1}$ and terminated at field strengths greater than ≈ 8T [35], when the energy of the first Landau level exceeds the Fermi energy.

The results of resistance measurements performed in the magnetic quantum limit (*H* > 8 T) are shown in Fig. 2 for a structure of thin graphite with a single nanohole produced by FIB (Fig. 2a) and a graphene structure with a single nanohole produced by the helium ion microscope (Fig. 2b). The common feature of both structures was the presence of field-periodic oscillations at high magnetic fields.

We compared the oscillation period for three single-nanohole samples of different diameters (see Table 1). Within the experimental uncertainty of approximately 10%, we found that the oscillation period Δ*H* for all samples corresponded to the flux quantisation in a hole

$$\Delta H \, \pi D^2 / 4 = \Phi_0 \qquad (1)$$

where $\Phi_0 = hc/e$ is the flux quantum and *D* is the diameter of the hole. This result is expected if one considers that the main contribution to the effect comes from the carriers with orbits that are localised very near the edge of the hole. Such a periodicity was observed for the first time in ring-shaped samples of graphene by S. Russo et al. [27]. We can therefore attribute the oscillations to quantum interference of massless DFs in a band of conducting edge states.

An interesting question is why the orbits of the DFs that contribute to the oscillations are so close to the edge. Actually, there are conventional skipping cyclotron orbits of DFs around the nanohole. However, they should not exhibit the interference effect. The reason is that skipping orbits circle around the hole in the same direction independent of valley index. In contrast, TD states in two valleys revolve in opposite directions. This leads to emergence of resonances in intervalley backscattering, which will be explained below, see section "Theory and comparison with experiment". Therefore, only TD states that exist close to the nanohole edge may cause the oscillations.

TD-type edge states can play the role of a ring that keeps carriers near the edge of the hole. We can estimate the effective width of this type of effective ring. Table 1 presents a comparison of the geometrical size of the hole and the size of the effective 1D ring. One can see that the radius of the effective ring is always greater the geometrical radius and the difference is approximately 2 nm (within the experimental accuracy). This value gives an experimental estimate of the width of the effective ring that is associated with the edge states, i.e., the penetration depth of the edge states.

We now discuss the temperature dependence of the amplitude of the oscillations. For experiments with graphene rings, the AB oscillations have been observed only below the temperature range of liquid helium [35]. The effect on nanoholes persists to much higher temperatures. For example, Fig. 3a shows the oscillating component of the magnetoresistance extracted from the data of Fig. 2b by subtracting the monotonic part. One can see that oscillations persist up to temperatures as high as 50 K. The four main peaks marked by the upper arrows are clearly observed at field strengths of greater than 10 T, when the magnetic quantum limit is realised (i.e., the energy of the first Landau level exceeds the Fermi energy). Their spacing $\Delta H$ corresponds to the flux quantum per nanohole area, following Eq. 1.

Fig. 3b shows the temperature dependence of the height of one of these peaks $A$ observed at $H = 18\,\text{T}$. This figure clearly shows an exponential dependence, $A \propto \exp(T/T_0)$, with $T_0 \approx 17\,\text{K}$. Weak T-damping of oscillations is consistent with the theory of edge states that is discussed below.

The next important point revealed from the experiment is the existence of the relatively small peaks that are marked in Fig. 3a by upward arrows. The two series of peaks can be considered as shifted by $\pi$ series with the same $\Phi_0$ periodicity or as series of oscillations with periodicities $\Phi_0$ and $\Phi_0/2$. We can extract the temperature dependence of this type of oscillation only from the peak at $H = 10\,\text{T}$, which is more or less clearly resolved. A comparison of the temperature dependence of the peaks of these two different series demonstrates the same exponential dependence, thus indicating their common origin.

One can consider that the characteristic temperature $T_0$ is related to the typical energy of the edge state as $kT_0 = E_0 = \hbar v_0 / R$. This relationship gives an estimate for $v_0$, $v_0 = 5 \cdot 10^6\,\text{cm/s}$. To compare the temperature dependences for samples #2 and #3, we found that for nanoholes of smaller diameter, the $T_0$ value increases approximately proportional to $1/R_{eff}$.

## Theory and comparison with experiment

The band structure of graphene consists of two almost-independent cones, which are often called valleys. Let us colour them red and blue for convenience, as indicated in Fig. 4. The electronic structure of graphene is modified near an edge such that TD-type edge states can appear. The wave function of carriers at the edge state is exponentially localised near the interface. Neglecting the inter-valley scattering, one can characterise the edge [23, 24] by a real phenomenological parameter $a$. This parameter is included in the boundary conditions for the Weyl-Dirac equations describing the massless DFs. In fact, real atomic structure of graphene edge around the hole is unknown. In frames of our phenomenological approach [23] the edge parameter $a$ should depend on the tangential DF coordinate along the edge, but we simulate the edge by an average value of this parameter. This approximation corresponds to an average dynamic of the edge states circling around the hole. The value of $a$ can be determined by comparison with experiment. However, a comparison with the microscopic model calculations [24] indicates that the value is small: $|a| << 1$. Below, we will use this smallness.

For a half-plane graphene sample, the spectrum of the edge states in a zero magnetic field is [23, 24]:

$$E_\tau(k_\parallel) = \tau 2 a v \hbar k_\parallel, \qquad \tau k_\parallel \geq 0$$

(see Fig. 4a). Here, $k_\parallel$ is the one-dimensional electron momentum along the edge measured from the centres of the valleys, $v = 10^8$ cm/s is the effective speed of light in the bulk graphene spectrum, and the index $\tau = \pm 1$ enumerates the valleys in graphene. For small $a$, the localisation length of the edge state at the Fermi level is $x_T = 1/|k_r| = 1/|k_b|$. A comparison with experiment, which is presented in Table 1, yields the estimate $x_T = 2$ nm.

Edge states also exist at the edges of a round hole ("antidot") of radius $R$ in an infinite sheet of graphene. DFs trapped in the edge states behave like electrons in a narrow ring ($x_T \ll R$), which causes us to expect manifestations of the Aharonov-Bohm effect. The finite perimeter of the antidot leads to a quantisation of the tangential component of the momentum. The discrete edge states are thus characterised by half-integer total angular momenta $j_\tau$ ($\tau j_\tau > 0$). In the quasiclassical approach, the spectrum of these edge states is obtained from (2) by substituting $k_\parallel$ for $j_\tau / R$. Although these states are quasistationary in the absence of a magnetic field, their finite lifetime (due to their decay into the bulk states) is large in the actual case of small $a$.

In a magnetic field $H$, the spectrum of the TD edge states in an antidot has quasiclassical asymptotes at $|a| \ll 1$:

$$E_\tau = \tau 2a \frac{\hbar v}{R}(j_\tau + \Phi/\Phi_0 - \tau/2) \quad (3)$$

where $\Phi = H\pi R^2$, $\Phi_0 = hc/e$ and the half-integer $j_\tau$ satisfies the condition $\tau(j_\tau + \Phi/\Phi_0) > 0$.

Under the conditions of our experiment, the asymptotes given by (3) are valid. Knowing $x_T = 2$ nm and the value of the field corresponding to the last Shubnikov-de Haas oscillation, we can estimate from Eq. (1) the absolute value of the parameter $a$ to be $|a|=0.05$.

Bulk DFs move on trajectories in smooth random potential (Fig.4d). If the trajectory is close enough to the antidot, bulk DFs can tunnel to the edge state at the antidot. Then a fermion in the edge state moves periodically around the antidot (clockwise in one valley and counterclockwise in another), thereby acquiring additional Aharonov-Bohm phase in a magnetic field. Its wave function satisfies the Bloch theorem. The spectrum of the edge states has a band character, and the magnetic flux plays the role of an effective quasi-momentum. Intra-valley scattering does not essentially change this picture. Consider weak inter-valley scattering, which potentially acts as a periodic potential. This leads to the formation of energy gaps in the band spectrum in Fig. 4b if the following conditions of inter-valley ("blue-red") resonance are fulfilled:

$$\frac{\Phi}{\Phi_0} = \frac{j_+ + j_-}{2} \quad (4)$$

These resonances lead to a strong backscattering, which, in turn, according to the Landauer formula, results in the negative peaks in the conductance. Consider the magnetic quantum limit, when the AB oscillations are observed. Then, the Fermi level is close to the Dirac point, which fluctuates strongly in space because of the formation of "puddles" of electrons and holes [43]. If the spatial scale of these fluctuations is comparable to the size of the antidot, and if the energy scale is comparable to the energy of the perimetric quantisation of the edge states, the resonance (4) does not depend on the position of the Fermi energy and its temperature smearing. This qualitatively explains the weak temperature dependence of the observed AB oscillations.

As demonstrated above, the velocity of the edge Dirac fermions extracted from the experimental temperature dependence of the amplitude of oscillations is 20 times less than that of the bulk fermions. This difference may be explained by considering the smallness of the edge parameter, $|a|=0.05$. This value is consistent with the independent estimate of $a$ made above.

With the magnetic field variation, the energies of the TD states in the red valley $E_+(\Phi, j_+)$ periodically coincide with the TD energies in the blue valley $E_-(\Phi, j_-)$ with $\Phi_0/2$ periodicity. Because $j$ takes only half-integer values, it follows from (4) that the flux $\Phi/\Phi_0$ can accept either integer or half-integer values. The first condition is responsible for the main series of oscillations, whereas the second one results in the complementary series.

## Conclusions

In summary, we found that graphene-on-graphite and graphene nanostructures that contain single nanoholes exhibit field-periodic resistance oscillations with magnetic flux periodicity that is approximately equal to the flux quantum per nanohole area. This result is considered to be the Aharonov-Bohm effect due to conducting states localised near the edge of the hole. Such states, which are called Tamm-Dirac states, are manifestation of Tamm-Shockley edge states in two-dimensional systems of massless Dirac fermions. From the experiment, we have obtained an estimate of the values of the penetration depth and the velocity of these states. The proposed mechanism of the oscillations is based on resonant inter-valley back-scattering of the Tamm-Dirac states.

Recently, a nanohole array in nanoperforated graphene was studied using gate voltage spectroscopy in the absence of a magnetic field [44]. Discrete levels that are presumably related to the edge states were observed. However, the most important distinctive feature of the Tamm-Shockley states -- the ability to conduct electrical current – has not been directly demonstrated. Moreover, the effects of the mutual influence of neighbouring nanoholes in the array do not have an unambiguous interpretation. Therefore, the question of the origin of the detected levels remained open.

In this study, we obtained the first direct evidence for band conduction of Tamm-Shockley-type edge states. In graphene, the Tamm-Dirac edge states are populated by massless Dirac fermions, the low temperature conductivity of which is metallic. The conducting properties of these non-topological states are similar to the properties of topological edge states in topological insulators that are known in the literature. This analogy is apparently connected to the similarity of the edge spectra of massless Dirac fermions in these two different systems.

## Acknowledgments

The work has been supported partly by RFBR and the associated International Laboratory between the Neel Institute (France) and the Kotelnikov Institute (Russia).

## References

1. Shur, M. *Physics of Semiconductor Devices* (Prentice Hall, New Jersey, 1990).
2. Tamm, I. E. Uber eine mogliche Art der Elektronenbindung an Kristalloberflachen. *Phys. Z. Sowjetunion* **1**, 733–736 (1932).
3. Shockley, W. On the surface states associated with a periodic potential. *Phys. Rev*. **56**, 317–323 (1939).
4. Duke, C. B. *Surface Science: The First Thirty Years* [C.B. Duke (ed.)] (North-Holland, Amsterdam, 1994).
5. Joannopoulos, J. D., Johnson, S. G., Winn, J. N., Meade, R. D. *Photonic Crystals: Molding The Flow Of Light* (Princeton Univ. Press, New Jersey, 1995).
6. Vinogradov, A. P., Dorofeenko, A.V., Merzlikin, A. M., Lisyansky, A. A. Surface states in photonic crystals. *Phys. Usp.* **53** 243–256 (2010).
7. Hasan, M. Z., Kane, C. L. Colloquium: Topological insulator. *Rev. Mod. Phys.* **82**, 3045 (2010).
8. Qi, X.L., Zhang, S.C. Topological insulators and superconductors. *Rev. Mod. Phys.* **83**, 1057 (2011).

9. Fu, L., Kane, C. L. Topological Insulators with Inversion Symmetry. *Phys. Rev. B*. **76**, 045302 (2007).
10. Chen, Y. L. et al. Experimental Realization of a Three-Dimensional Topological Insulator, $Bi_2Te_3$. *Science* **325**, 178 (2009).
11. Xia, Y. et al. Observation of a large-gap topological-insulator class with a single Dirac cone on the surface 2009, *Nature Phys*. **5**, 398 (2009).
12. Hsieh, D. et al. Observation of Unconventional Quantum Spin Textures in Topological Insulators. *Science* **323**, 919 (2009).
13. Peng, H. et al. Aharonov–Bohm interference in topological insulator nanoribbons. *Nat. Mater.* **9**, 225 (2010).
14. Xiu, F. et al. Manipulating surface states in topological insulator nanoribbons. *Nat. Nanotechnol.* **6**, 216 (2011).
15. Bardarson, J. H., Moore, J. E. Quantum interference and Aharonov–Bohm oscillations in topological insulators. *Rep. Prog. Phys.* **76**, 056501 (2013).
16. Volkov, V. A., Pinsker, T. N. Spin splitting of the electron spectrum in finite crystals having the relativistic band structures. *Sov. Phys. Solid State* **23**, 1022 (1981).
17. Volkov, B. A., Pankratov, O. A. Two-dimensional massless electrons in an inverted contact. *JETP Lett.* **42**, 178 (1985).
18. Kusmartsev, F. V., Tsvelik, A. M. Semimetallic properties of a heterojunction, *JETP Lett.* **42**, 257 (1985).
19. Jackiw, R., Rebbi, C. Solitons with fermion number 1/2. *Phys. Rev. D* **13**, 3398 (1976).
20. Geim, A. K., Novoselov, K. S. The rise of graphene. *Nat. Mater.* **6**, 183 (2007).
21. Nakada, K., Fujita, M., Dresselhaus, M. S. The edge state in graphene ribbons: Nanometer size effects and edge shape dependence. *Phys. Rev. B*. **54**, 17954-17961 (1996).
22. Castro Neto, A. H., Guinea, F., Peres, N. M. R., Novoselov, K. S., Geim, A. K. The electronic properties of graphene. *Rev. Mod. Phys.* **81**, 109–162 (2009).
23. Volkov, V. A., Zagorodnev, I. V. Electrons near a graphene edge. *Low Temp. Phys.* **35**, 2-5 (2009).
24. van Ostaay, J. A. M., Akhmerov, A. R., Beenakker, C. W. J., and Wimmer, M. Dirac boundary condition at the reconstructed zigzag edge of graphene. *Phys. Rev. B* **84**, 195434 (2011).
25. Ritter, K. A., and Lyding, J. W. The influence of edge structure on the electronic properties of graphene quantum dots and nanoribbons. *Nat. Mater.* **8**, 235-242 (2009).
26. Tao, C. et al. Spatially resolving edge states of chiral graphene nanoribbons. *Nature Phys.* **7**, 616-620 (2011).
27. Russo, S. et al. Observation of Aharonov-Bohm conductance oscillations in a graphene ring. *Phys. Rev. B*. **77**, 235404 (2007).
28. Huefner, M. et al. Investigation of the Aharonov-Bohm effect in a gated graphene ring. *Phys. Status Solidi B*, **245**, 2756-2759 (2009).
29. Wurm, J., Wimmer, M., Baranger, H. U., and Richter, K. Graphene rings in magnetic fields: Aharonov-Bohm effect and valley splitting. *Semicond. Sci. Technol.* **25**, 034003 (2010).
30. Huefner, M. et al. The Aharonov-Bohm effect in a side gated graphene ring. *New. J. Phys.* **12**, 043054 (2010).
31. Weng, L., Zhang, L., Chen, Y. P., and Rokhinson, L. P. Atomic force microscope local oxidation nanolithography of graphene. *Appl. Phys. Lett.* **93**, 093107 (2008).
32. Yoo, J. S., Park, Y. W., Skakalova, V., and Roth. Shubnikov-de Haas and Aharonov Bohm effects in a graphene nanoring structure. *Appl. Phys. Lett.* **96**, 143112 (2010).
33. Smirnov, D., Schmidt, H., and Haug, R. G. Aharonov-Bohm effect in an electron-hole graphene ring system. *Appl. Phys. Lett.* **100**, 203114 (2012).
34. Nam, Y. et al. The Aharonov-Bohm effect in graphene rings with metal mirrors. *Carbon.* **50**, 5562-5568 (2012).
35. *for review see* Schelter, J., Recher, P., Trauzettel, B. The Aharonov-Bohm effect in graphene rings. *Sol. State Commun.* **152**, 1411-1419 (2012).

36. Shen, T. et al. Magnetoconductance oscillations in graphene antidot array. *Appl. Phys. Lett*. **93**, 122102 (2008).
37. de Juan, F., Cortijo, A., Vozmediano, M. A. H., Cano, A. Aharonov-Bohm interferences from local deformations in graphene. *Nature Phys.* **7**, 810 (2011).
38. Li, G., Luicann, N., and Andrei, E. Y. Scanning Tunneling Spectroscopy of Graphene on Graphite. *Phys. Rev. Lett.* **102,** 176804 (2009).
39. Neugebauer, P., Orlita, M., Faugeras, C., Barra, A.-L, and Potemski, M. How Perfect Can Graphene Be? *Phys. Rev. Lett.* **103**, 136403 (2009).
40. Faugeras, C. et al. Magneto-Raman Scattering of Graphene on Graphite: Electron Scattering and Phonon Excitations. *Phys. Rev. Lett.* **107**, 036807 (2011).
41. Kuhne, M. et al. Polarization-resolved magneto-Raman scattering of graphene-like domains on natural graphite. *Phys. Rev. B*. **85**, 195406 (2012).
42. Latyshev, Yu. I. et al. Graphene production by etching natural graphite single crystals in a plasma-chemical reactor based on beam-plasma discharge. *Doklady Physics*. **57**, 1-3 (2012).
43. Das Sarma, S., Adam, S., Hwang, E., Rossi, E. Electronic transport in two-dimensional graphene. *Rev. Mod. Phys.* **83**, 407 (2011).
44. Latyshev, Yu. I. et al. Orbital quantization in a system of edge Dirac fermions in nanoperforated graphene. *JETP Lett*. **98**, 214 (2013).

**Author contribution statement**

Yu.I.L. and V.A.V. devised the project. Yu.I.L. and A.P.O. designed and fabricated the single-hole nanostructures with FIB. O.F.V. and Yu.V.P. fabricated the single-hole nanostructures with the helium ion microscope. Yu.I.L. and A.P.O. performed measurements and analysed the results. P.M. supported the experiments performed at the High Magnetic Field Laboratory in Grenoble. V.A.V., I.V.Z., and V.V.E. provided theoretical support. Yu.I.L. and V.A.V. wrote the paper. All authors contributed to discussions.

**Additional Information**

Competing financial interests.
The authors declare no competing financial interests.

**Table 1.**

The table includes the following samples: #1 – graphene structure with a hole produced by HIM, #2 – thin graphite structure with a hole produced by FIB, and #3 – thin graphite structure with a hole produced by HIM. The thicknesses of the thin graphite structures were varied in the range of 30-50 nm. The parameter $D_{eff}$ was calculated using Eq. (1).

| Sample No | $\Delta H$, T | $D_{geom.}$, nm | $D_{eff.}$, nm | $(D_{geom} - D_{eff})/2$, nm |
|---|---|---|---|---|
| #1 | 9.0 | 20±1 | 24±0.1 | 2.0±0.5 |
| #2 | 3.2 | 37±2 | 41± 0.2 | 2.0±1.0 |
| #3 | 6.0 | 25±1 | 30±0.2 | 2.5±0.5 |

**Figures**

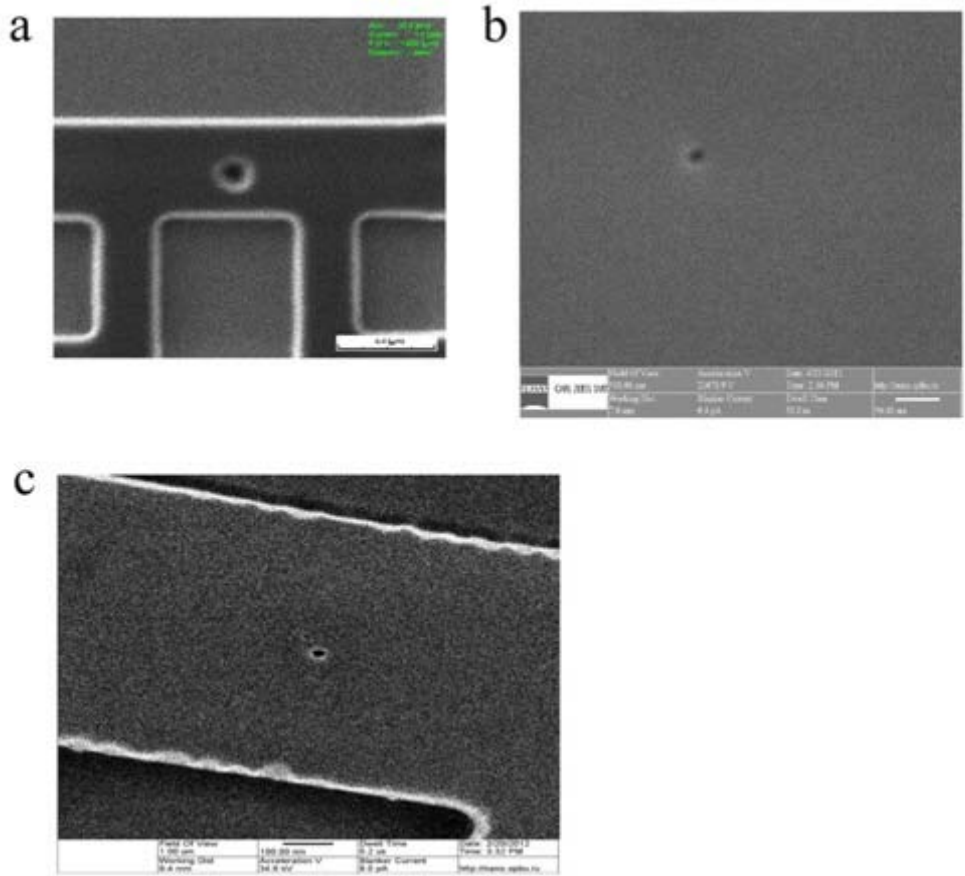

*Figure 1. Experimental realisation of graphene nanohole structures.*

*Single hole produced by FIB (SEM image, **a**) and by the helium ion microscope (SHIM image, **b, c**) in graphene (**b**) and thin graphite (**a, c**).*

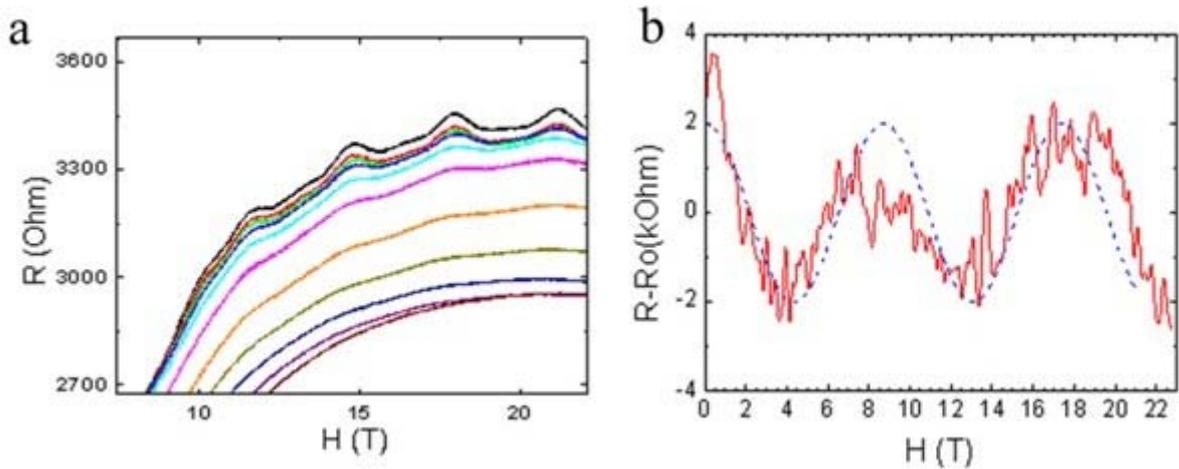

*Figure 2. Aharonov-Bohm resistance magneto-oscillations.*

*Field-periodic resistance oscillations for **a**) a thin graphite single nanohole (D=37 nm) structure made by FIB at various temperatures: T=1.5, 4.2, 10, 15, 20, 30, 45K (from top to bottom) and **b**) a graphene structure with a single nanohole made using a helium ion microscope, $D = 20$ nm, T=4.2 K.*

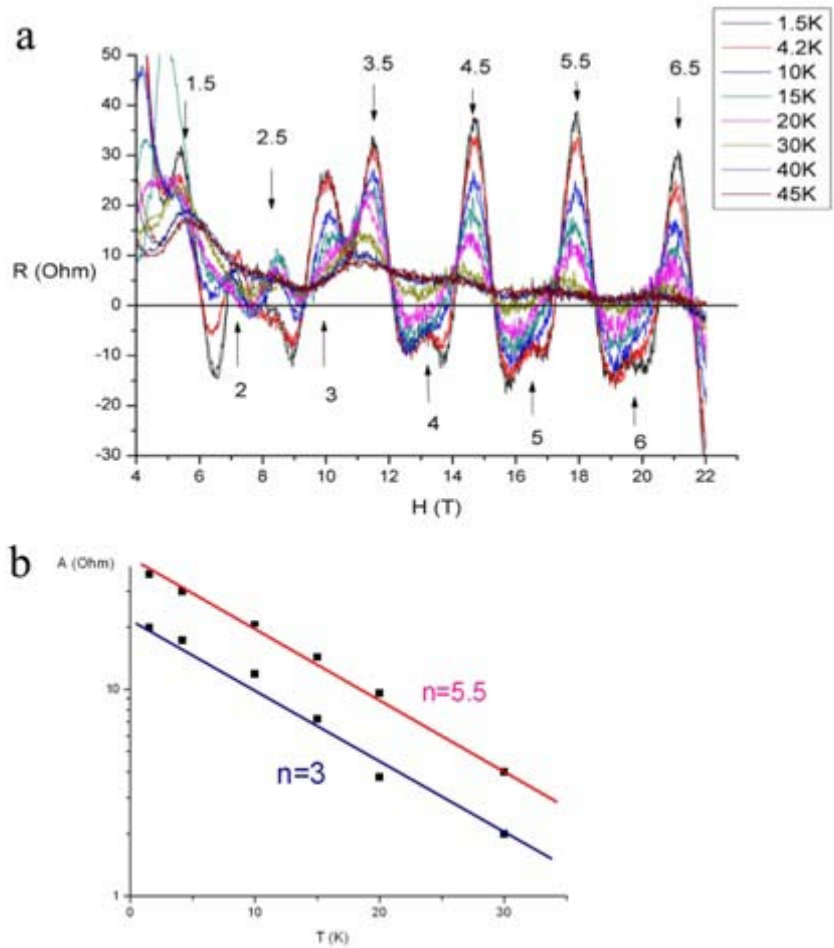

***Figure 3. Temperature behaviour of magneto-oscillations for sample #2.***
*a)  Oscillating part of the resistance at various temperatures. The downward arrows indicate the main series, which corresponds to $\Phi_\downarrow = n\Phi_0 + 1/2$ (where n is an integer), whereas the upward arrows mark an additional series $\Phi_\uparrow = n\Phi_0$. **b)** Temperature dependences of the oscillation amplitude for $\Phi/\Phi_0 = 5.5$ (red line) and 3 (blue line).*

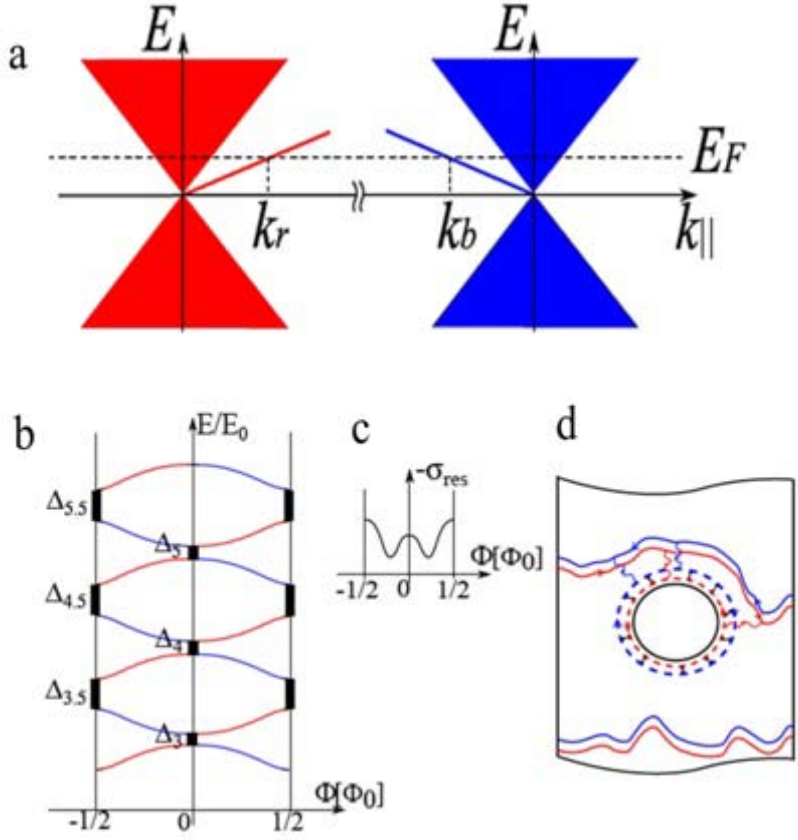

*Figure 4. Edge states around the graphene nanohole*

*a) The red and blue rays are the Tamm-Dirac states contra-propagating along the graphene semi-plane. There are two Tamm-Dirac states at the Fermi level: $k_r$ in the red valley with positive velocity and $k_b$ in the blue valley with negative velocity. The bulk continuum states are shadowed.*

*b) Spectrum of the Tamm-Dirac states in an antidot for different $j_+$, $j_-$ as a function of the magnetic flux that passes through the antidot area. The spectrum has a band character, and the flow through the antidot plays a role of quasi-momentum in the reduced zone scheme. The red (blue) colour corresponds to the valley with $\tau = +1$ ($\tau = -1$), $E_0 \approx 2a\hbar v/R$. The red-blue scattering results in band gaps (vertical bold lines). Gaps are formed by anticrossing of red and blue edge states with angular momenta $j_+$ and $j_-$. Gap values are denoted by the index $j = (j_+ + j_-)/2$.*

*c) Inter-valley contribution to the conductivity in the reduced zone scheme. The peaks correspond to resonant red-blue back-scattering. The two series of peaks are connected by the passing of the magnetic flux through the centre and boundary of the zone shown in panel b.*

*d) Contra–propagating trajectories of the orbit centres for different valleys for the zero Landau level (N=0) in a smooth-impurity potential. One of the orbits is close to the antidot and can experience inter-valley back-scattering at the above-mentioned values of the flux.*